\documentclass[conference]{IEEEtran}
\IEEEoverridecommandlockouts
\pdfoutput=1
\usepackage{cite}
\usepackage{amsmath,amssymb,amsfonts}
\usepackage{algorithmic}
\usepackage{graphicx}
\usepackage{textcomp}
\usepackage{xcolor}
\usepackage[utf8]{inputenc}
\def\BibTeX{{\rm B\kern-.05em{\sc i\kern-.025em b}\kern-.08em
    T\kern-.1667em\lower.7ex\hbox{E}\kern-.125emX}}
\begin{document}

\title{Mutual Influences in Interwoven Systems and their detection in the context of Organic Computing}

\author{\IEEEauthorblockN{Neeraj Mumbuveetil Sasankan}
\IEEEauthorblockA{\textit{Intelligent Technical Systems} \\\textit{University of Passau} \\
Passau, Germany \\
mumbuv01@gw.uni-passau.de}}

\maketitle

\begin{abstract}
Technical systems have evolved over time into large and complex Interwoven Systems consisting of several to a huge number of (possibly heterogeneous) subsystems that have interdependencies. The resultant mutual influences among subsystems have made them so complex that they are no longer manageable by humans and it is assumed to intensify rapidly. Identifying such mutual influences is the first step towards mastering the complexity of such systems. This paper presents mutual influences in Interwoven Systems by describing real-world examples and a methodology to detect them in the context of Organic Computing. The methodology is evaluated with the help of an example. Further, a taxonomy of Organic Computing applications helpful for selecting suitable methods for detecting hidden mutual influences is described briefly.
\end{abstract}


\section{Introduction}
 
Technical systems have become an unavoidable part of life and they form an important part of the environment that supports us in our daily life. They help in improving efficiency in the tasks carried out in environments ranging from industries to households. Various challenges are involved in the lifetime(i.e., design, development, integration, deployment, maintenance) of such a system. The complexity of such systems increases rapidly as they grow in size to meet the changing needs of their application field. This results in systems that consist of a large number of subsystems that interact with one another. The challenges involved in integrating such subsystems reaches a point when they can no longer be anticipated at design time. Organic Computing (OC) initiative [1] uses the concepts of self-adaptation and self-organisation to master the complexities involved. This resulted in systems with several to a huge number of entities with interdependencies called Interwoven Systems [2], [3]. Such interdependencies are referred to as mutual influences here. Mutual influences can be explicit or implicit. Explicit mutual influences are easy to observe but implicit mutual influences do not reveal easily to the observer. The awareness of such mutual influences is necessary for maintaining an optimal system behaviour [3].

Consider {\it SmartGrid} as an example of how a traditional system can grow into a system which shows an interwoven structure with new hidden mutual influences as described in [4]. Power systems used to be strictly centralised and pre-planned. The rise of renewable energy (such as biogas, solar and wind), the introduction of electric vehicles, the possibility to control demand (e.g., due to smart meters), and similar developments have triggered a dramatic change in the overall energy system [5]. Such systems, which were formerly governed by the principle of "separation of concerns",  now exhibit an interwoven system structure. The number of independently operating power plants belonging to a variety of operational authorities has increased dramatically. This is accompanied by a direct and indirect coupling via the previously unconsidered communication network [6]. Specific challenges include simultaneously charging electric vehicles and the potential effects of price-based incentives to change consumption policies. In such a complex system, identifying mutual influences is very important for maintaining an optimal behaviour. This article presents a methodology to detect mutual influences and a taxonomy of Organic Computing systems helpful for selecting suitable methods for detecting hidden mutual influences.

The remainder of this article is organised as follows. Section II describes the utilized system model which is inspired by standard machine learning notions. Section III gives a methodology to detect hidden mutual influences and subsequently, an evaluation of the methodology using a smart camera network is described in section IV. This is followed by a taxonomy of Organic Computing systems helpful for selecting suitable methods for detecting hidden mutual influences in section V. The section VI that follows, explains the taxonomy with a smart camera network as an example. Finally, the article is concluded in section VII.

\section{System Model}

Consider a system model as described in [4]. The system consists of a set of agents \{$A_1${$,\dots,$}$A_n$\}. Each agent has the ability to assume different configurations. Let $A_i$ be the $i^{th}$ agent, the configuration space of $A_i$ is given by $C_i$ $=$ {$c_{i1}$}{$\times$$\cdots$$\times$}{$c_{im}$}, where $c_{ij}$ are the parts of the configuration. The configurations of individual agents are assumed to be non-overlapping. This means that each agent has its own set of configurations, $c_{ij}$ $\neq$ $c_{kl}$ for all {$i \neq k,j,l$}. But the configuration parts do not have to be completely disjoint in structure and values of the contained variables. For example, consider two routers who might have the possibility to configure the time-out interval. This will lead to the same set of possible configurations in these attributes on different devices. Such a relation is explicitly allowed within the model.

Apart from configuration space, a further element to be considered is the {\it local performance measurement}. In order to apply the proposed method, each agent has to estimate the success of its decisions at runtime - as a response to actions taken before. This is realized based on a feedback mechanism. The feedback can be {\it direct}, in which case the feedback possibly originates from the environment of the agent, or {\it indirect}, in which case the feedback is manually assigned.

\section{Methodology}

The methodology described here for the measurement of mutual influences has been originally presented in [7]. The objective is to identify those parts of the configuration of the neighbouring agents that have an influence on the agent itself. Although the focus is on spacially neighbouring agents, the methodology is also applicable to virtually neighbouring agents (e.g., routers in a data communication network). After the influencing configuration parts have been identified, they can be addressed by a designer or by a self-adapting system itself. The interest is more on the detection of specific configuration parameters whose optimal usage strategy is somehow influenced by the current settings of the neighbouring agents. The basic idea of the algorithm is to make use of stochastic dependency measures that estimate associations and relations between the configuration parts of an agent and the performance of a second agent. These dependency measures are designed to find correlations between two random variables {\it X} and {\it Y}. The {\it Maximal Information Coefficient} (MIC - [8]) is the dependency measure used here.

To understand the Maximal Information Coefficient, {\it Mutual Information} [9] must be understood first. The mutual information is defined as: 
\begin{equation}
{I(X;Y)} = \sum_{x\in X} \sum_{y\in Y} p(x,y) \log{\left(\frac{p(x,y)}{p(x)p(y)}\right)}
\end{equation}
where $p(x,y)$ is the joint probability distribution of the discrete random $X$ and $Y$ variables. In addition, $p(x)$ and $p(y)$ are the corresponding marginal distributions. The measure quantifies how much information about $X$ can be retrieved from the realization of $Y$ and vice versa. The mutual information gives values $\geq 0$ and only equals to zero if the two random variables are completely stochastically independent. An important advantage of this technique is the possibility to find non-linear dependencies. The probability distributions $p(x)$, $p(y)$, and $p(x,y)$ are unknown and have to be estimated in order to calculate the mutual information. In the discrete case, this is mostly done by counting the frequency of occurrence of different events. For continuous random variables, the formula is given by: 
\begin{equation}
I(X;Y) = \int_{x\in X}\int_{y\in Y} p(x,y) \log{\left(\frac{p(x,y)}{p(x)p(y)}\right)} dxdy
\end{equation}
In this case also, $I(X;Y)\geq 0$ is 0 if and only if both the random variables are completely stochastically independent. Also, it can be used to find non-linear dependencies. But a drawback in this estimation process is that it is not possible to use the straightforward method of counting frequencies for continuous variables.

Having understood mutual information, MIC is explained in this section. MIC is based on mutual information. It uses {\it binning} of samples to overcome the drawback in the continuous case. The data is sorted into bins based on their similarity following which the probability distributions are estimated for the bins. This is essentially a discretization of the data. Using the resulting distributions, the discrete variant of the mutual information is calculated. The problem is that the manual choice of the bins is time-consuming and can lead to deceptive results if not appropriate. Therefore, MIC has a concept of always using bins that lead to the maximal mutual information. As finding this bin configuration is computational heavy, MIC utilizes a heuristic to tackle the problem. As a result, MIC is defined as: 
\begin{equation}
MIC(X;Y) = \max_{n_x n_y < B} {\frac{I(X;Y)}{\log(\min(n_x,n_y))}}
\end{equation}
where $n_x$ and $n_y$ denote the number of bins for $X$ and $Y$. The divisor, $\log(\min(n_x,n_y))$, is used as a normalising factor as it gives the maximal achievable mutual information given the number of bins. $B$ denotes a function of the sample size $N$ and limits the number of bins. This is necessary to avoid {\it trivial} partitioning, such as creating a single bin for each data point that most of the time result in relatively high values for the mutual information. The initial paper introducing MIC proposes to use $B = N^{0.6}$.

In general, MIC shows some interesting properties. It is defined for values $\geq 0$ and is equal to zero only if the random variables are completely independent. Besides, it is normalised and shows a good {\it equitability} in the simulation results.

This method does not take into account the configuration of the agent itself which leads to some issues. These issues are explained in the following example. Consider two agents $A$ and $B$, each of which can take two configurations $C_1$ and $C_2$. A configuration consists of only one attribute which can take only $1$ or $0$ as its value. The performance of agent $B$ with respect to the configurations of the two agents is defined in Tab. I [7]. It can be figured out easily that the performance of $B$ is high (i.e., $1$) if it takes on the same configuration as $A$, and low (i.e., $0.5$) if it takes on a configuration different from that of $A$. Intuitively, it can be stated that $A$ influences $B$. The actual calculation is done as follows. The calculations are done using mutual information for simplicity. However, the issue applies to other dependency measures (such as MIC) as well. The calculation requires, the probabilities of different performances of $B$, denoted as $p_B(x)$, the probabilities of different configurations of $A$, denoted as $p_A(y)$, and the joint probability of both, denoted as $p_{BA}(x,y)$. It is assumed that the agents take on one of the two configurations with probability $0.5$. 
\begin{table}[htbp] 
\caption{THE PERFORMANCE OF AGENT $B$ GIVEN THE DIFFERENT
CONFIGURATIONS OF THE TWO AGENTS}
\begin{center}
\begin{tabular}{|c|c|c|}
\hline
\textit{$P_B$}&\textit{$c_1$}&\textit{$c_2$} \\
\hline
\hline 
\textit{$c_1$} & $1$ & $0.5$ \\
\hline
\textit{$c_2$} & $0.5$ & $1$ \\
\hline
\end{tabular}
\label{tab1}
\end{center}
\end{table}
Given this, the values of the needed probabilities can be calculated from Tab. I. as $p_B(1) = p_B(0.5) = p_A(c_1) = p_A(c_2 ) = 0.5$ and $p_{BA}(1,c_1) = p_{BA}(1,c_2) = p_{BA}(0.5,c_1) = p_{BA}(0.5,c_2) = 0.25$. Thus, the resulting mutual information is:  $I(X;Y)$ $=$ $\sum_{x}$ $\sum_{y} p_{BA}(x,y)$ $\log{\left(\frac{p_{BA}(x,y)}{p_B(x)p_A(y)}\right)} =$  $\sum_{x}$  $\sum_{y} 0.25$ $\log{\left(\frac{0.25}{0.5\cdot 0.5}\right)} =$ $\sum_{x\in X}$ $\sum_{y\in Y} 0.25\cdot 0$ $= 0$ which indicates that $A$ does not influence $B$. This is unsatisfying for the context of the purpose of the technique.

This can be resolved by taking into consideration not only the configuration of the {\it remote agent} (here $A$) but also the configuration of the agent itself (here $B$). This is because the influence is equal for every configuration of $A$ but only appears for one configuration of $B$. This effect can be avoided by calculating the dependency of the performance and the configuration for each configuration of $B$ separately. But, for agents with large configuration spaces, this solution becomes infeasible since the sample size decreases linearly with the number of configurations. Hence, too much time would be required to detect the influences. Therefore, the configuration is split into two parts and dependency is calculated separately. Therefore, the issue is resolved and the sample size for each calculation remains the highest.

\section{Evaluation}

This section gives a brief idea of how to proceed with the methodology in an example scenario of a smart camera network. In the original work detailed in [7], the mutual influence between a set of cameras is measured which is then used for designing a learning algorithm. What is presented here are certain aspects to be considered when applying the methodology in the given scenario. 

\subsection{Smart Camera Networks}

The term {\it Smart Camera} means a camera equipped with a built-in computation unit (which can be utilized for tasks, such as image processing, object localization and object tracking). Also, most Smart Cameras have pan, tilt and zoom (PTZ) capabilities and the computation unit is used to determine beneficial alignments for the camera. Beyond, Smart Cameras are equipped with wired or wireless communication devices that allow for communication with neighbouring cameras and a command and control centre. 

The evaluation described here is an extension of the general model for smart cameras, as provided in [10], to meet the requirements of the mutual influence scenario. Every camera in the system has a location $(x_l, y_l, z_l)\in \mathbb{R}^3$ with $z_l\gg 0$ and the scene is modelled as a plane $D\subset \mathbb{R}^3$ with the normal form $0x_p+0y_p+z_p=0$, with $(x_p, y_p, z_p)\in \mathbb{R}^3$. The performance of the cameras and the whole system is measured as follows. The goal is to observe {\it things}, i.e., the interest is in what happens at a specific place at some specific point in time (the existence of sophisticated image processing algorithms is assumed). Therefore, the performance is measured with respect to {\it interestingness I} which is modelled as a function depending on the place $(x,y)\in \mathbb{R}^2$. Let $I_t$ be defined as $I_t : D\rightarrow [0,\infty) \subset \mathbb{R}, (x,y) \mapsto I_t(x,y)$, for every time $t\in \mathbb{N}$. The higher the value of $I_t(x,y)$ at time $t$, the more interesting the point $(x,y)$ is. If $I_t(x,y)=0$, the point $(x,y)$ is not interesting at all at time $t$. To determine the performance of a camera $c$, the observed points $V_{ct} \subset D$ of this camera based on its PTZ configuration being active in the time step $t$ has to be taken into account. To calculate the performance $P_{ct}$ of camera $c$ at time $t$, the equation, $P_{ct} = \int_{V_{ct}} I_t(x,y)dxdy$ can be used. This holds for a scenario with a single camera. In a scenario where there are multiple cameras, the possibility that two or more cameras observe the same spot has to be considered. This may be useful in some situations (for instance, stereo- reconfiguration of suspicious persons) but may not be useful in some other situations (for instance, observe as much space as possible). In order to address this issue, a factor $n_t$ is applied to modify the previous formula as $P_{ct} := \int_{V_{ct}} I_t(x,y)n_tdxdy$. The sum of performances of individual cameras will give the performance of the whole system. For the measurement of the performance according to the model, some specific functions $I_t(x,y)$ and $n_t$ are required. These are chosen based on the application scenario. For instance, if the focus is on the detection of interesting targets, $I_t(x,y)$ is $1$, if undetected target on $(x,y)$ and $0$, if there is nothing new to observe on $(x,y)$. Since the camera's field of view potentially covers more than just one location, $n_t = {\frac{1}{m}}$ where $m$ is the number of observing cameras.

The experimental setup used and the results of the experiment for two different scenarios are detailed in [7]. The results show that mutual influence detection, using the proposed methodology, works well.

\section{Taxonomy}

This section gives a taxonomy of Organic Computing systems helpful for selecting suitable methods for detecting hidden mutual influences as detailed in [11]. These important characteristics of OC systems can be used to give a guideline on how to measure the influences in the system.   

\subsection{Agents}

The number of agents in a system is an interesting characteristic for the influence detection. By observing OC systems, the following categories can be identified based on the number of agents:

\begin{itemize}
\item Small systems, i.e., systems with a few agents.
\item Middle-size systems, i.e., systems with less than a few hundred agents.
\item Large-scale systems, i.e., systems with more than a few hundred agents.
\end{itemize}

Another characteristic regarding the agents is the configuration space of them. As described before in the system model in Section II, a system can have multiple configuration parts that can have different forms. The number of configuration parts is an interesting criterion. Furthermore, the following types of configuration parts can be identified:

\begin{itemize}
\item Nominal: the different values can be categorized, but there is no order for the categories. For example, categories like left and right.
\item Ordinal: the categories can be ordered. For example, categories like low, medium and high, or 1, 2, 3.
\item Infinite real-valued: an infinite number of values can be assumed. For example, this could be an interval [0,1].
\end{itemize}

For the types nominal and ordinal, there is although another characteristic for classification which is the number of categories. In contrast, for the infinite real-valued class, it is always assumed that the set of values is infinite.

As an extension to this section described in [11], the type of agents in terms of the internal hardware components should also be considered. Although agents might appear homogeneous to the outside, because of the difference in the internal components they will have to be considered as heterogeneous agents. This could also be an important aspect to consider in the detection of mutual influences.

\subsection{Communication}

Different system types are possible based on the communication possible between the agents and the associated cost. In the context of detection of mutual influences, there are two border cases. The first one is the case where communication is free with all agents in the system. This can happen if the system is composed of virtual agents that utilize the same hardware which would lead to negotiable communication cost. The second border case is that the communication is strictly limited to the neighbours. This becomes a border case since limitation to no communication at all makes no sense as a potential influence cannot be detected. Between these two cases, there is a variety of possibilities that reach from low to high costs for multi-hop communication.

\subsection{Influence}

In the context of high communication costs, one characteristic that is particularly interesting is that the influence could originate from an entity that is in the neighbourhood or one that can only be contacted over multiple hops. This leads to different detection possibilities based on the possible communication.

Another characteristic of the influences is the possibility that multiple agents have to act jointly in order to reveal the mutual influence between them. For example, consider a scenario in which two robotic arms are required to hold a workpiece together while the other drills a hole in it. In this case, observing the robotic arms separately does not reveal any influence.

The strength of influence is also an important characteristic. There are two aspects to this classification. The first is the type of dependency based on the power of the dependency measures. Some of them are limited to linear or monotonic dependencies, but the more powerful measures can measure stochastic dependencies which is the most general class of dependencies. The second aspect is the strength of the reflection of dependency in the joint distribution; it can be very distinctive or rather not distinctive.

Temporal aspect is another important aspect to be considered. In some cases, the results are reflected right after the configuration is assumed, meaning that the influence is immediate. However, there are also cases where the results appear after a delay.

\section{Example for taxonomy}

This section gives an example for the depicted taxonomy as described in [11]. Here, the smart camera networks depicted previously is classified based on the taxonomy. Starting with the entity characteristics. Considering the first characteristic, that is the number of entities, there is no clear classification since SCN can have different sizes from a few cameras to a few hundred cameras. Large-scale systems are also possible. The number of configuration parts in this domain is three, i.e., the pan tilt and the zoom. Each of the configuration parts is infinite real-valued.

Regarding the communication, there are different instances of SCN. As mentioned before, the connection between smart cameras can be wired or wireless. In the case of a wireless ad-hoc network, the communication cost is high at least for multi-hop communication. In the case of wired connections, the cost is not as high as in wireless, but, it has its limitations.

The influences in such a system are limited to the spatial neighbourhood. This is because the cameras can only be influenced by other cameras that share the potential field of view due to the nature of the performance measure. But it cannot be assumed that the influences can be detected by linear or monotonic measures. Therefore, they are categorized as stochastic. Moreover, no instance can be found in which the influence only reveals if several neighbours act in common. Furthermore, a temporal influence is not found in the system since the previous configurations of a camera do not affect other cameras.

\section{Conclusion}

To conclude the work, this article presented the importance of mutual influences in Interwoven Systems in the context of Organic Computing. Further, a methodology to detect hidden mutual influences is presented in detail which is explained with a smart camera network as an example. Finally, a taxonomy of Organic Computing systems helpful for selecting suitable methods for detecting hidden mutual influences is presented with an extension to the existing aspects.

Overall, the article serves as a guide for the detection of hidden mutual influences not only by describing an effective methodology but also by laying out the important characteristics of OC systems that will aid in the same.


\clearpage

\begin{thebibliography}{00}

\bibitem{b1} C. Müller-Schloer, H. Schmeck, and T. Ungerer, Eds., Organic Computing - A Paradigm Shift for Complex Systems. Birkhäuser, 2011. 

\bibitem{b2} K. L. Bellman, S. Tomforde, and R. P. Würtz, “Interwoven systems: Self-improving systems integration,” in Eighth IEEE International Conference on Self-Adaptive and Self-Organizing Systems Workshops, SASOW 2014, London, United Kingdom, September 8-12, 2014, 2014, pp. 123–127.

\bibitem{b3} S. Tomforde, J. Hähner, H. Seebach, W. Reif, B. Sick, A. Wacker, and I. Scholtes, “Engineering and Mastering Interwoven Systems,” in Proc. of ARCS 2014 Workshops, 2014, pp. 1–8.

\bibitem{b4} S. Tomforde, S. Rudolph, K. L. Bellman, and R. P. Würtz, “An organic computing perspective on self-improving system interweaving at runtime,” in 2016 IEEE International Conference on Autonomic Computing, ICAC 2016, Wuerzburg, Germany, July 17-22, 2016, 2016, pp. 276–284. DOI : 10.1109/ICAC.2016.15.

\bibitem{b5} J.-P. Steghöfer, G. Anders, F. Siefert, and W. Reif, “A System of Systems Approach to the Evolutionary Transformation of Power Management Systems,” in Informatik 2013, 43. Jahrestagung der Gesellschaft für Informatik e.V. (GI), Informatik angepasst an Mensch, Organisation und Umwelt, 16.-20. September 2013, Koblenz, ser. LNI, vol. 220, 2013, pp. 1500–1515.

\bibitem{b6} Appelrath, Hans-Jürgen, Henning Kagermann, and Christoph Mayer. "Future energy grid." Migrationspfade ins Internet der Energie. acatech Studie. Deutsche Akademie der Technikwissenschaften (2012).

\bibitem{b7} S. Rudolph, S. Tomforde, B. Sick and J. Hähner, "A Mutual Influence Detection Algorithm for Systems with Local Performance Measurement," 2015 IEEE 9th International Conference on Self-Adaptive and Self-Organizing Systems, Cambridge, MA, 2015, pp. 144-149.

\bibitem{b8} Reshef, David N., Yakir A. Reshef, Hilary K. Finucane, Sharon R. Grossman, Gilean McVean, Peter J. Turnbaugh, Eric S. Lander, Michael Mitzenmacher, and Pardis C. Sabeti. "Detecting novel associations in large data sets." science 334, no. 6062 (2011): 1518-1524.

\bibitem{b9} C. Shannon and W. Weaver, The Mathematical Theory of Communication. University of Illinois Press, 1949.

\bibitem{b10} Rudolph, Stefan, Sarah Edenhofer, Sven Tomforde, and Jörg Hähner. "Reinforcement learning for coverage optimization through PTZ camera alignment in highly dynamic environments." In Proceedings of the International Conference on Distributed Smart Cameras, p. 19. ACM, 2014.

\bibitem{b11} Rudolph, Stefan, and Sven Tomforde. "A taxonomy for organic computing systems regarding mutual influences." (2016).

\end{thebibliography}
\end{document}